\newcommand{\CLASSINPUTbottomtextmargin}{1in}
\documentclass[conference]{IEEEtran}
\usepackage{graphics,psfrag}
\usepackage{algorithm}
\usepackage{algpseudocode}
\usepackage{amsbsy}
\usepackage{url}
\usepackage[hang,small]{caption}
\usepackage{subcaption}
\usepackage[mathscr]{euscript}
\usepackage{color}
\usepackage[usenames,dvipsnames]{xcolor}

\usepackage{graphicx}
\ifCLASSINFOpdf
\else
\fi
\usepackage[cmex10]{amsmath}

\begin{document}
%
\title{Multi-Metric Energy Efficient Routing in Mobile Ad-Hoc Networks}

\author{\IEEEauthorblockN{Evripidis Paraskevas\IEEEauthorrefmark{1},
Kyriakos Manousakis\IEEEauthorrefmark{2},
Subir Das\IEEEauthorrefmark{2}  and
John S. Baras\IEEEauthorrefmark{1}}
\IEEEauthorblockA{\IEEEauthorrefmark{1}Institute for Systems Research and Department of Electrical \& Computer Eng\\
University Of Maryland, College Park, Maryland 20742 \\ \IEEEauthorrefmark{2}Applied Communication Sciences, Basking Ridge, NJ 07920 \\ Email: \{evripar, baras\}@umd.edu, \{kmanousakis, sdas\}@appcomsci.com}}


\maketitle

\begin{abstract}
Increasing network lifetime by reducing energy consumption across the network is one of the major concerns while designing routing protocols for Mobile Ad-Hoc Networks. In this paper, we investigate the main reasons that lead to energy depletion and we introduce appropriate routing metrics in the routing decision scheme to mitigate their effect and increase the network lifetime. For our routing scheme, we take into consideration multiple layer parameters, such as MAC queue utilization, node degree and residual energy. We integrate our multi-metric routing scheme into OLSR, a standard MANET proactive routing protocol. We evaluate via simulations in NS3 the protocol modifications under a range of different static and mobile scenarios. The main observations are that in static and low mobility scenarios our modified routing protocol leads to a significant increase (5\%-20\%) in network lifetime compared to standard OLSR and slightly better performance in terms of Packet Delivery Ratio (PDR).

\textit{Keywords}-Routing, Energy efficient, OLSR, MANET
\end{abstract}


%
\IEEEpeerreviewmaketitle

\section{Introduction}
Mobile communication systems without central management have been gaining popularity in the form of Mobile Ad-Hoc Networks (MANETs). MANETs have widespread applications ranging from military scenarios, handling emergencies and natural disasters to distributed processing of data. This type of mobile networks do not rely on any pre-established infrastructure and all nodes should be able to communicate with other nodes directly or indirectly through intermediate nodes. Hence, routing is a key operation in MANET and the appropriate selection of routing scheme affects its performance. For this reason designing efficient routing schemes has been an extensive research area in the last decade. Many proactive (e.g. OLSR~\cite{995315}) and reactive routing protocols (e.g. AODV~\cite{Perkins:2003:AHO:RFC3561}) for MANET have been developed. 

Limited battery life of mobile nodes impose an important limitation in the performance of this type of networks. Power depletion of a mobile node affects its ability to forward packets on behalf of other nodes and may lead to partitioning of the network.  Therefore, efficient battery utilization and increase of network lifetime should be important design criteria for developing routing schemes in MANET. 

While many energy-efficient routing protocols have been presented (\cite{Yu03energyefficient, Singh:1998:PRM:288235.288286,Jones:2001:SEE:500484.500486}) to optimize energy consumption across the network, these protocols are mostly based on a single routing metric, which is derived from energy measurements. For example, some of the current techniques use the reciprocal value of the residual energy to do minimum energy consumption routing (\cite{ee-olsr,Kunz}) or they use Minimum Drain Rate (MDR) mechanism (\cite{ee-olsr,novel-DSR}). These approaches, based on a single metric, are myopic and do not consider all the possible causes of energy depletion in the network. Thus, the investigation of a combination of network parameters (multiple metrics), which are not strictly derived by residual energy but still indicate energy depletion levels in parts of the network, will contribute to designing more efficient routing schemes.

Our insight for this work is to determine a wider set of cross-layer parameters, not solely based on energy measurements, that enable effective prediction of low energy paths, while encouraging uniform utilization of network resources. First, we identify the reasons that cause energy depletion in different parts of the network and then choose our metrics to mitigate their effect. We take into consideration three metrics in our routing scheme: \textit{MAC queue utilization}, \textit{residual energy} and \textit{node degree}. \textit{MAC queue utilization} is introduced to help our routing scheme to predict and avoid congested parts of the network, which are subject to high energy consumption due to the number of packet transmissions. \textit{Residual energy} is crucial, because we want to choose paths, which include less depleted nodes. Finally, \textit{node degree} contributes to reducing the energy consumption due to overhearing in the neighbor nodes of a possible intermediate node of the selected path.

In this work, we focus on designing a novel multiple metric routing scheme for MANET, based on the described metrics, and integrate it to standard OLSR to examine its effectiveness. We combine the multiple metrics to compute a weight for each node. These weights are efficiently propagated to the rest of the network. Then, we propose a weight-based routing scheme, which will use a greedy approach to choose the path with the lower cost, computed from the weights the nodes received. In this way, the routing scheme takes into account the multiple metrics introduced and switches paths in order to avoid energy-depleted, highly congested and dense areas of the network. The contributions of this work are:
\begin{itemize}
\item introduction of a combination of routing metrics from multiple layers, which has not been examined in any previous work related to energy efficient routing;
\item proposal of a novel multi-metric routing scheme, which takes into account the above metrics, and integrate it to standard OLSR;
\item experimentation with the modified OLSR using NS3~\cite{NS3} and examination of its energy behavior and its performance (in terms of Network Lifetime and PDR) compared to the standard OLSR
\end{itemize}

The rest of the paper is organized as follows. In section~\ref{sec:related_work} we shortly describe the prior work for energy efficient routing. In section~\ref{sec:OLSR} we briefly introduce OLSR. We describe our multi-metric routing scheme in
section~\ref{sec:multi metric}. We verify our claims by extensive simulations in section~\ref{sec:simulation}.

\section{Related Work}\label{sec:related_work}
The problem of designing energy-efficient routing protocols has received significant attention by the research community for over a decade. Many energy-efficient routing schemes have been developed (\cite{Yu03energyefficient, Singh:1998:PRM:288235.288286,Jones:2001:SEE:500484.500486}), which are typically based on residual energy derived metrics. In particular, a lot of research has been conducted on modifying standard routing protocols in MANET, such as AODV (\cite{Jones:2001:SEE:500484.500486, Gao}), DSR(\cite{novel-DSR}) and OLSR (\cite{ee-olsr, Kunz, DeRango:2009:EEO:1688291.1688320, Mahfoudh, Guo}).

Many energy-efficient variations of OLSR modify both the MPR selection and the route computation algorithm. For the MPR selection, some protocols~\cite{Kunz} choose the 1-hop neighbors with the maximum residual energy, while some others modify the willingness metric of the 1-hop neighbors (\cite{ee-olsr, DeRango:2009:EEO:1688291.1688320}) based on the energy level. In addition, an algorithm that takes into account the residual energy of the 1-hop and 2-hop neighbors of the MPR candidate is presented in~\cite{Mahfoudh}. To make the routing decision, most of the techniques proposed above modify the routing metrics to take into account energy consumption. Commonly used metrics for computing path costs, such as reciprocal value of residual energy (\cite{ee-olsr, Kunz}) and drain rate in intermediate nodes (\cite{ee-olsr,novel-DSR}), are only based on energy measurements.

Multiple metrics routing schemes, which are more related to our work, have also been proposed. An adaptive multiple metrics routing scheme for AODV is introduced in~\cite{Gao}. The authors take into account three routing metrics, which are hop count, traffic load and energy cost, and combine them to evaluate the cost of the paths. In addition, a predictive multiple metrics routing scheme for proactive protocols, like OLSR, is presented in~\cite{Guo}. The chosen routing metrics are mean queueing delay, energy cost and residual link lifetime. The authors designed a multi-objective routing schemes that evaluates the multiple metrics in a composite way and achieves better performance in terms of Packet Delivery Ratio (PDR) and network lifetime. 

\section{Brief Review of OLSR}\label{sec:OLSR}
Optimized Link State Protocol (OLSR)~\cite{995315} is an optimization of link state routing protocol, which inherits the stability of a traditional link state algorithm and adds the advantage of its proactive routing nature to provide routes immediately when needed. In OLSR, like in all proactive routing protocols, the nodes periodically broadcast control packets (HELLO and topology control packets (TC)) to find their 1-hop neighbors and advertise a subset of their links. Upon receipt of these packets, each node calculates and updates routes to each known destination. The key concept of reducing the overhead in OLSR is the multipoint relays (MPRs) ~\cite{Qayyum}. The multipoint relays of a mobile node are nodes in its 1-hop neighborhood that are selected in order to forward the node's topology control (TC) packets. Every node maintains a set of these nodes which constitutes its MPR Selectors set. In addition, every node chooses to advertise only the neighbors that are in its MPR selectors set in the topology control (TC) packets that it broadcasts. Thus, MPRs are used as intermediate nodes to form a route from a given node to any destination. 

MPRs of a given node are selected based on some criteria. First, they should have a bi-directional link with the node. Second, the MPR set should cover all nodes that are in the 2-hop neighborhood of the given node. The protocol is designed to work in a distributed way without the need of a central entity. In addition, it performs hop by hop routing, which means that each node has the information for the next hop towards the destination in its routing table and it forwards the packet accordingly. Finally, OLSR has good performance in large and dense networks.

\section{Multi-Metric Energy Efficient Routing Scheme for the OLSR protocol}\label{sec:multi metric}
The main goal of our routing scheme is to increase the network lifetime, without loss of performance, and we use OLSR as a case study. For the purpose of this work, lifetime is defined as the time until the battery of \textit{any} mobile node of our ad-hoc network depletes. We adopt this definition because, in the worst case, the depletion of a node may possibly cause network partition. To prevent the energy depletion and increase the network lifetime, we need to take into account cross layer parameters. These parameters include network congestion, residual energy of mobile nodes, as well as, network topology parameters. We aim to modify OLSR to make routing decisions according to these parameters and measure the performance improvement of our approach compared with the standard OLSR, using various performance metrics.

In this section, we introduce the routing metrics for our scheme and describe the modifications that need to be done in the OLSR protocol. We present the metrics propagation mechanism in detail and a greedy approach to make routing decisions based on the current values of the metrics.
\subsection{Routing Metrics}
We are mainly interested in defining the network parameters that affect the network lifetime, which will be our main performance metric. An intuitive way to select an appropriate set of metrics for energy-efficient routing is to first investigate the causes of energy depletion. The residual energy in mobile nodes is being depleted in two ways~\cite{ee-olsr}:
\begin{enumerate}
\item \textbf{Packet Transmission:} each transmission causes energy consumption at the mobile node
\item \textbf{Overhearing from the neighbor nodes:} Due to the broadcast nature of the wireless channel, all the nodes in the neighborhood of a sender node may overhear its packets transmission, even if they are not the receivers. Reception of these packets results to unnecessary expenditure of battery energy of the recipients.
\end{enumerate}

The proposed routing scheme takes into account three routing metrics to estimate the path cost and make the routing decision:
\begin{enumerate}
\item \textbf{MAC queue utilization}: This parameter indicates network congestion. When a mobile node has to transmit a lot of packets then this will lead to a significant energy consumption. Thus, larger weight should be assigned to nodes with high MAC queue utilization.
\item \textbf{Residual energy}: This parameter is crucial in order to determine the next-hop node. The traffic should be directed to nodes that have enough residual energy to transmit. Hence, we should assign a large weight to nodes that have small residual energy to do forwarding.
\item \textbf{Node Degree}: The degree of a node is the number of nodes that belong to its one-hop neighborhood. As we mentioned before, one reason for energy depletion is overhearing. We will try to avoid forwarding packets through nodes with high degree, because this will cause greater overall energy depletion. In addition, lower degree nodes also reduce the size of the interference graph, so fewer collisions will happen during our packet transmissions.
\end{enumerate}

We use a weight-based routing scheme, where a weight is assigned dynamically at each node. Mobile nodes update their routing tables according to the path costs computed using the nodes' weights received at each time period. The metrics are normalized by their maximum values and they contribute additively to the node's weight computation with some multiplicative factors, as shown in Equation~\ref{eq:metric}.

\begin{equation}
\label{eq:metric}
w_{i}=\alpha_1 \frac{L_{i}}{L_{max}} + \alpha_2 (1-\frac{E_{i}}{E_{max}}) + \alpha_3 \frac{D_{i}}{D_{max}} ,
\end{equation}
where $\sum\limits_{i=1}^3 \alpha_{i}=1$, $L_{i}$ is the number of packets in the MAC queue, $E_{i}$ is the residual energy at each time and $D_{i}$ is the node degree. $L_{max}$ is the maximum considered MAC queue size, $E_{max}$ is the initial energy of a node and $D_{max}$ is the number of nodes in the network minus one.
 By varying the weighting factors, we can change the importance of the three routing metrics during route discovery. For our experiments, we gave equal value to all $\alpha_{i}$.

\subsection{Modifications in the Standard OLSR}
\subsubsection{\textbf{TC packet format}}
\begin{figure}
\centering
\includegraphics[width=.34\textwidth,keepaspectratio=true]{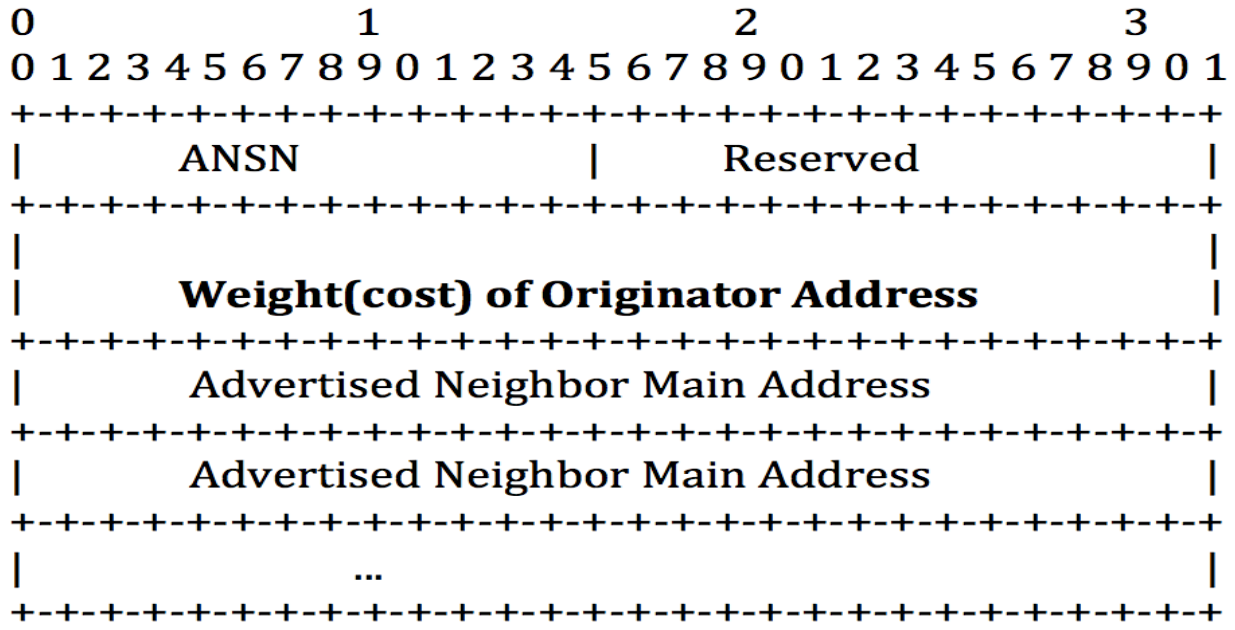}
\caption{TC packet format}
\label{fig:TC}
\end{figure}

The proactive nature of OLSR indicates that nodes need to periodically receive other nodes  weights, in order to compute the path costs and update their routing tables. Thus, it is crucial to find a way to propagate nodes weights to the network without increasing network overhead. An effective way is to embed this information to the TC packets that are periodically generated by each node. This does not introduce significant overhead, in contrast to new packets being generated to advertise the weights, and is easily implementable in routing protocols. Therefore, we extend the TC packet to include a field for the updated weight (computed locally using equation (1)) of the originator node, as shown in Figure~\ref{fig:TC}. Nodes receive and process the new TC packets and create topology tuples in their topology table. Topology tuples include the originator node of TC,  the destination node, which is one of the nodes advertised in the TC packet, and also the weight of the originator node.

\subsubsection{\textbf{MPR selection}}
We have not modified the way that OLSR selects MPRs in our scheme. It is based on the heuristic that the MPR set should cover all the nodes that are two hops away (two-hop neighborhood).
\subsubsection{\textbf{Routing Table Computation}}
OLSR is based on minimum hop routing to proactively determine the next-hop for a specific destination node. Thus, we need to change the way the routing table is computed using OLSR. The update of routing tables should be based on the path costs computed from the nodes weights and not on the number of hops. In addition, routing tables should include path costs to the destination address instead of the number of hops, as in standard OLSR. We define path cost as the sum of weights of the intermediate nodes along the path plus a constant weight, equal to 1, that corresponds to the source node. We do not consider the receiver's weight, because the energy consumption during transmission of a packet from an intermediate node is much greater than the power consumption of the receiver. To periodically update the routing table based on the computed path costs, we use the greedy heuristic scheme outlined in Algorithm~\ref{Routing Table Computation}.

\begin{algorithm}
\caption{Greedy Heuristic Algorithm}\label{Routing Table Computation}
\begin{algorithmic}[1]
\Procedure{RoutingTableComputation}{}
\State Clear Routing Table ()
\For{$i \gets 1, NumberofNeighbors$}
   \State Add New Entry in the Routing Table
   \State $cost \gets 1$
 \EndFor \\
 \For{$i \gets 1, NumberofTopologyTuples$} 
  \State $have\_last \gets Lookup(RouteTable, last\_addr)$ 
  \State $have\_dest \gets Lookup(RouteTable, dest\_addr)$ 
   \If {$have\_last \gets true \wedge have\_dest \gets false$} 
   \State Add New Entry in the Routing Table: 
   \State $dest \gets dest\_addr$ 
   \State  $next\_addr \gets last\_addr.next\_address$
   \State $interface \gets last\_addr.interface$
   \State $cost \gets last\_addr.cost + tc\_weight$ 
   \ElsIf {$have\_last \gets true \wedge have\_dest \gets true$} 
   \State UPDATE\_ROUTING\_ENTRY(last\_addr, dest\_addr, tc\_cost) \Comment If Needed
   \Else 
   \State Not adding Routing Table Entry 
   \EndIf
 \EndFor

\EndProcedure
\end{algorithmic}
\end{algorithm}
The algorithm assigns cost equal to 1 to paths towards the 1-hop neighbors. Then, it examines the topology tuples stored in the topology table and three cases are considered in order to update the routing table. The first one occurs when there is an entry in the routing table for the originator node (last\_addr) of the topology tuple. In this case, we add a new entry to the routing table for the destination node of TC (dest\_addr) with cost equal to the sum of the cost corresponding to route to the originator node (last\_addr.cost) and the originator node's weight. The second case, where the greedy nature of  our algorithm applies, occurs when there are entries for both the originator and the destination node of the topology tuple. Then, we choose greedily the new path detected through the originator node or we maintain the old path, by comparing their costs. Finally, in the case we do not have entries neither for the originator node nor the destination node, we do not create any new entry. 

An interesting result observed using the proposed modified routing scheme is that the nodes tend to choose longer paths, when the intermediate nodes of the currently selected paths have consumed a lot of energy. The observed path switching indicates that our routing scheme disperses the traffic across the network and utilizes the network resources in a more uniform way to increase network lifetime.

\section{Simulation}\label{sec:simulation}
In this section we will describe the different performance metrics used to evaluate our scheme, the simulation setup and the simulation results. For our simulations, we used the NS3 network simulator~\cite{NS3} and we created a modified version of OLSR to execute our experiments. 
\subsection{Performance Metrics}\label{sec:metrics}
The performance metrics that we consider for the evaluation of our modified scheme are:
\begin{enumerate}
\item \textbf{Packet Delivery Ratio (PDR) (in \%)}:  the ratio of the number of packets delivered to the destination nodes over the number of packets sent by the source nodes. We use the normalized PDR, which is defined as the number of packets delivered divided by the number of packets that should ideally been transmitted in this data rate. This gives more representative results.
\item \textbf{Network Lifetime (in sec)}: the time until the battery of a mobile node of our mobile ad-hoc network depletes.
\item \textbf{Average Node Residual Energy vs Time}: This metric was first proposed in~\cite{DeRango:2009:EEO:1688291.1688320}. It shows how the average energy consumption (total residual energy[J]/number of nodes) changes over time.
\item \textbf{Distribution of node residual energy}: This metric~\cite{Mahfoudh} indicates how the energy is consumed across the network. It shows how many nodes have the same percentage of residual energy at the end of the simulation. Ideally we should have a big number of nodes that will have 30-70 \% of their residual energy at the end of the simulation. This would illustrate that our modified scheme disperses the traffic across different paths in the mobile network and utilizes more uniformly the network resources. 
\end{enumerate}
\subsection{Simulation Setup}
We simulated a MANET with 30 nodes in a dense 2000 x 2000 meter square area, which has 4 hop network diameter. There are 3 CBR/UDP sources generating packets of 512 bytes with different data rates. To compute the average value of the metrics we want to derive, we simulated each scenario 3 times. In each different simulation, we choose different source-destination pairs. We have two variations of the simulation setup to evaluate the performance of our modified routing scheme compared to the standard OLSR. The common simulation parameters of the two variations are summarized in Table~\ref{tab:Simulation} below. 

\begin{center}

\captionof{table}{Common Simulation parameters}
\label{tab:Simulation}
\begin{tabular} {| c | c |}
\hline
    Area & 2000m x 2000m   \\ \hline
    Nodes & 30 \\ \hline
    Traffic Sources & 3 \\ \hline
    Traffic Type & CBR/UDP \\ \hline
    Packet Size & 512 bytes \\ \hline
    Start of Traffic & 30 sec \\ \hline
    Initial Node Energy & 7 Joules \\ \hline
    Transmission Power & 5 dbm \\ \hline
    Simulations/Scenario & 3 \\ \hline
    Link bandwidth & 1 Mbps \\ \hline
\end{tabular}
\end{center}

\subsubsection{\textbf{Setup A}}

In this setup, we want to evaluate the modified OLSR using the performance metrics that will give us clear view of the performance in terms of energy consumption. These metrics introduced in the previous section are: Average residual energy
vs time and Distribution of node residual energy. For this setup we consider a static scenario with the parameters described in Table I. In addition, the packet interarrival time is 0.1sec, which means that the source nodes send 10 packets/sec (around 41 Kbps). The simulation time is set to 250 sec.

\subsubsection{\textbf{Setup B}}

In this setup, we want to examine the performance of our modified OLSR compared to standard OLSR in a variety of static and mobile scenarios. These scenarios have the common parameters described in Table I. The two performance metrics that we will consider are Packet Delivery Ratio (PDR) and Network Lifetime (in sec). The simulations are done with 4 different packet interarrival times to study the effect of low, medium and high traffic rate in our scheme. The intervals are 0.1 sec (10 packets/sec or around 41 Kbps), 0.075 sec (14 packets/sec or around 55 Kbps), 0.05 sec (20 packets/sec or around 81 Kbps) and 0.025 sec (40 packets/sec or around 163 Kbps). At shorter intervals, a lot of packets will be lost due to network congestion and the bandwidth limitations, that we have set up for our experiments in NS3. Thus, we will notice a significant decrease in PDR as the interval becomes smaller. 
We also consider three different scenarios in terms of mobility: static, low mobility and high mobility. In the low mobility scenario, mobile nodes move in the area based on a Random Waypoint mobility model with maximum speed of 2 m/sec. In the high mobility scenario, nodes move based on a Random Waypoint mobility model with maximum speed of 20 m/sec. Finally, the simulation time is set to 150 sec for taking, under different data rates, comparative measurements for PDR. For network lifetime measurements, we execute the simulations until a node is completely depleted.

\subsection{Simulation Results}
In the following subsections, the results of our simulations in NS3 for Setup A and Setup B are presented.

\subsubsection{\textbf{Simulation Setup A}}
In this setup, we consider a static scenario with data rate of 10 packets/sec for our traffic flows. The purpose is to perform an energy analysis of the behavior of our modified scheme in comparison with standard OLSR. Hence, we examine the  two performance metrics described in the section~\ref{sec:metrics}.: Average Node Residual Energy vs Time and Distribution of node residual energy.

\begin{figure}
\centering
\includegraphics[width=.29\textwidth,keepaspectratio=true]{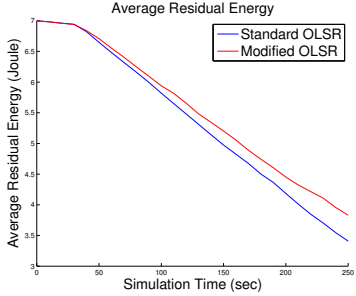}
\caption{Average Residual Energy in Simulation Setup A}
\label{fig:avgEnergy}
\end{figure}

For the first performance metric considered, we present the results in Figure~\ref{fig:avgEnergy}. The observation is that our modified scheme leads to a lower average power consumption than standard OLSR. We notice that there are 10-15\% energy savings by using the modified OLSR. 
 
\begin{figure}
\centering
\includegraphics[width=.29\textwidth,keepaspectratio=true]{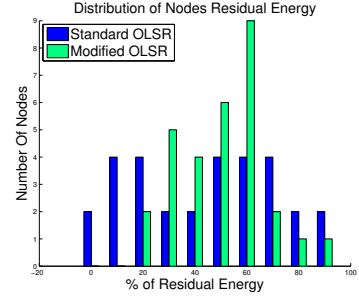}
\caption{Distribution of node residual energy in Setup A}
\label{fig:histogram}
\end{figure}

For the distribution of node residual energy, we present the results in Figure~\ref{fig:histogram}. We observe that in the modified OLSR than half of the nodes that have residual energy between 30-70\% at the end of the simulation time (250 sec).  On the other hand, in the standard OLSR there are a lot of nodes that have small percentage of residual energy or large percentage of residual energy. Thus, this indicates that our modified scheme achieves more uniform utilization of network resources by adjusting the weights and dispersing the traffic through different paths to reduce energy consumption. On the other hand, the standard OLSR selects the same paths to the destination nodes and utilizes the same intermediate nodes, which leads to fast depletion of their energy.

\subsubsection{\textbf{Simulation Setup B}}
In this setup, we consider static and mobile scenarios and also different data rates. The purpose of this experimentation is to examine whether the modified scheme contributes to the increase of network lifetime, but at the same time without loss of performance (in terms of PDR). The different mobility patterns indicate the behavior of the modified scheme under different network dynamics. 

\begin{figure}
        \centering
        \begin{subfigure}[b]{0.25\textwidth}
                \includegraphics[width=\textwidth]{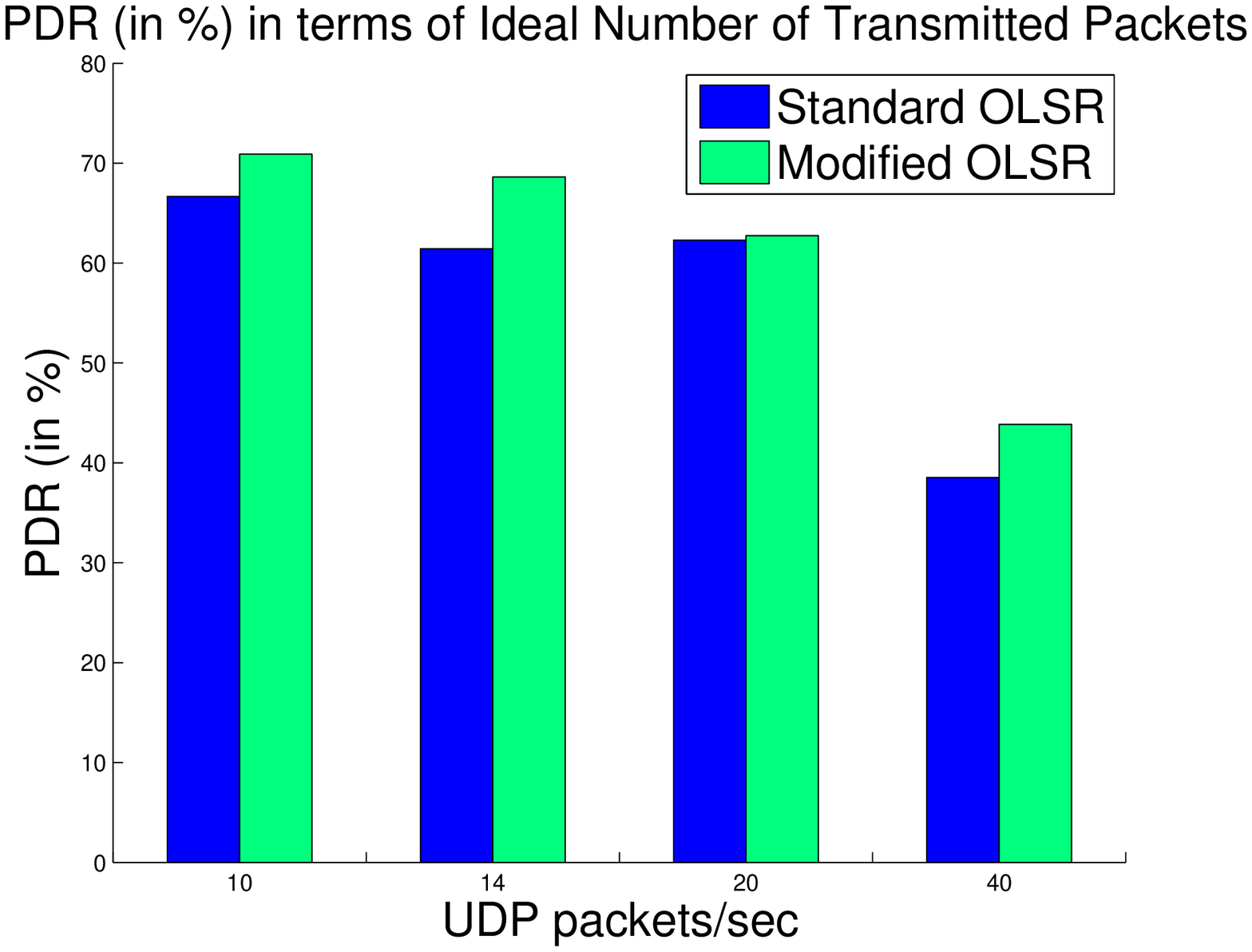}
                \caption{PDR for Various Packet Interarrival times}
                \label{fig:PDR Static}
        \end{subfigure}%
        ~ 
        \begin{subfigure}[b]{0.25\textwidth}
                \includegraphics[width=\textwidth]{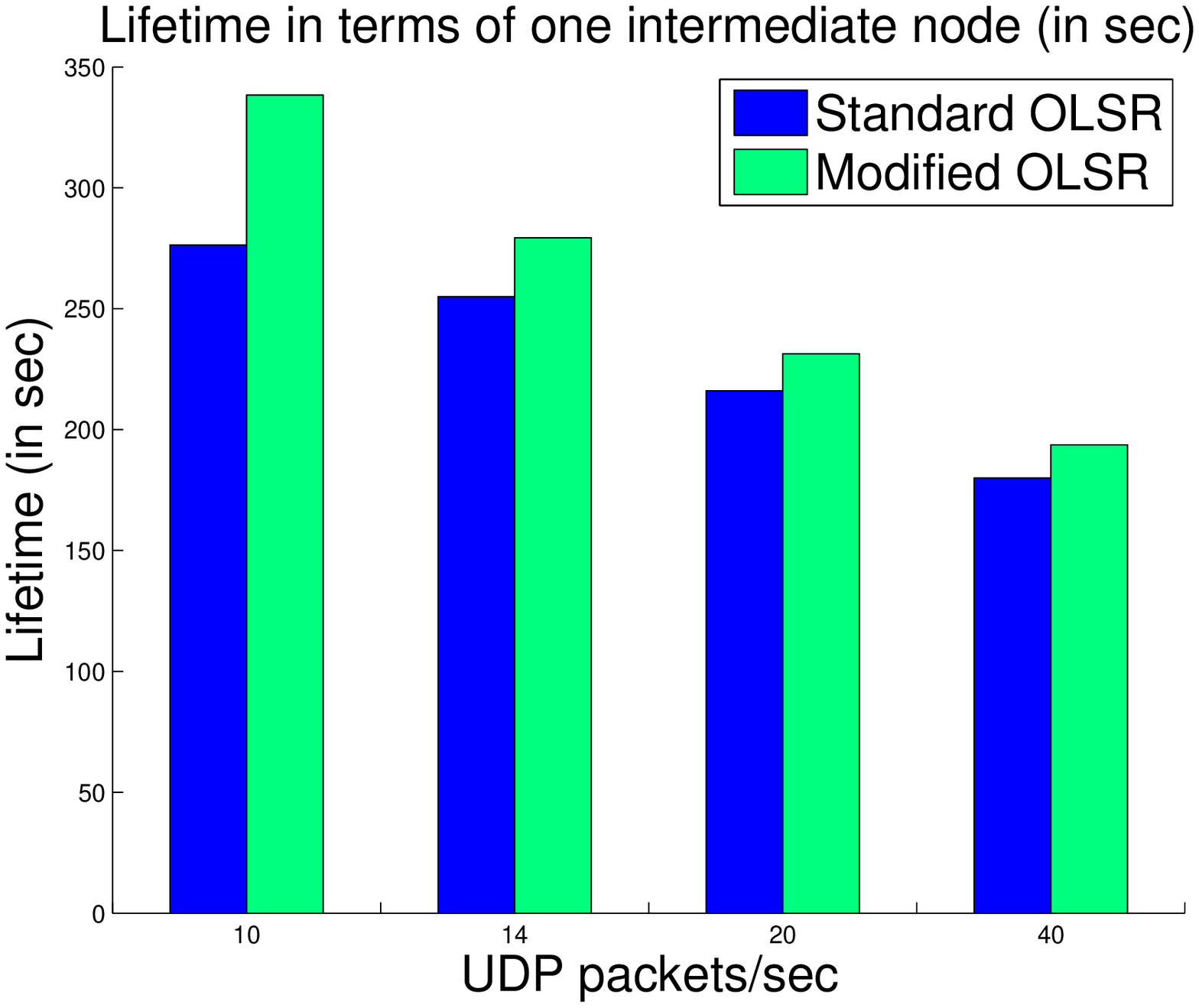}
                \caption{Network Lifetime for Vari-\\ous Packet Interarrival times}
                \label{fig:Network Lifetime Static}
        \end{subfigure}
  \caption{Static Scenario}
  \label{fig:Static Scenario}
\end{figure}

In the static scenario, we observe in Figure~\ref{fig:Network Lifetime Static} that the modified OLSR achieves improvement of 5-20\% in network lifetime, in comparison with the standard OLSR. This indicates that the modified scheme selects alternative paths, maybe longer paths, when an intermediate node has consumed most of its energy, and bypasses the energy "weak" node. In addition, PDR has an improvement of up to 10\% in the case of the modified OLSR (Figure~\ref{fig:PDR Static}). This is justified from the fact that queue utilization is taken into account as part of our weight function. Thus, the modified OLSR imposes higher weight to congested nodes, that are most likely to cause loss of packets, and avoids to choose to forward packets through them in the routing procedure. This causes this slight improvement in PDR.

\begin{figure}
        \centering
        \begin{subfigure}[b]{0.25\textwidth}
                \includegraphics[width=\textwidth]{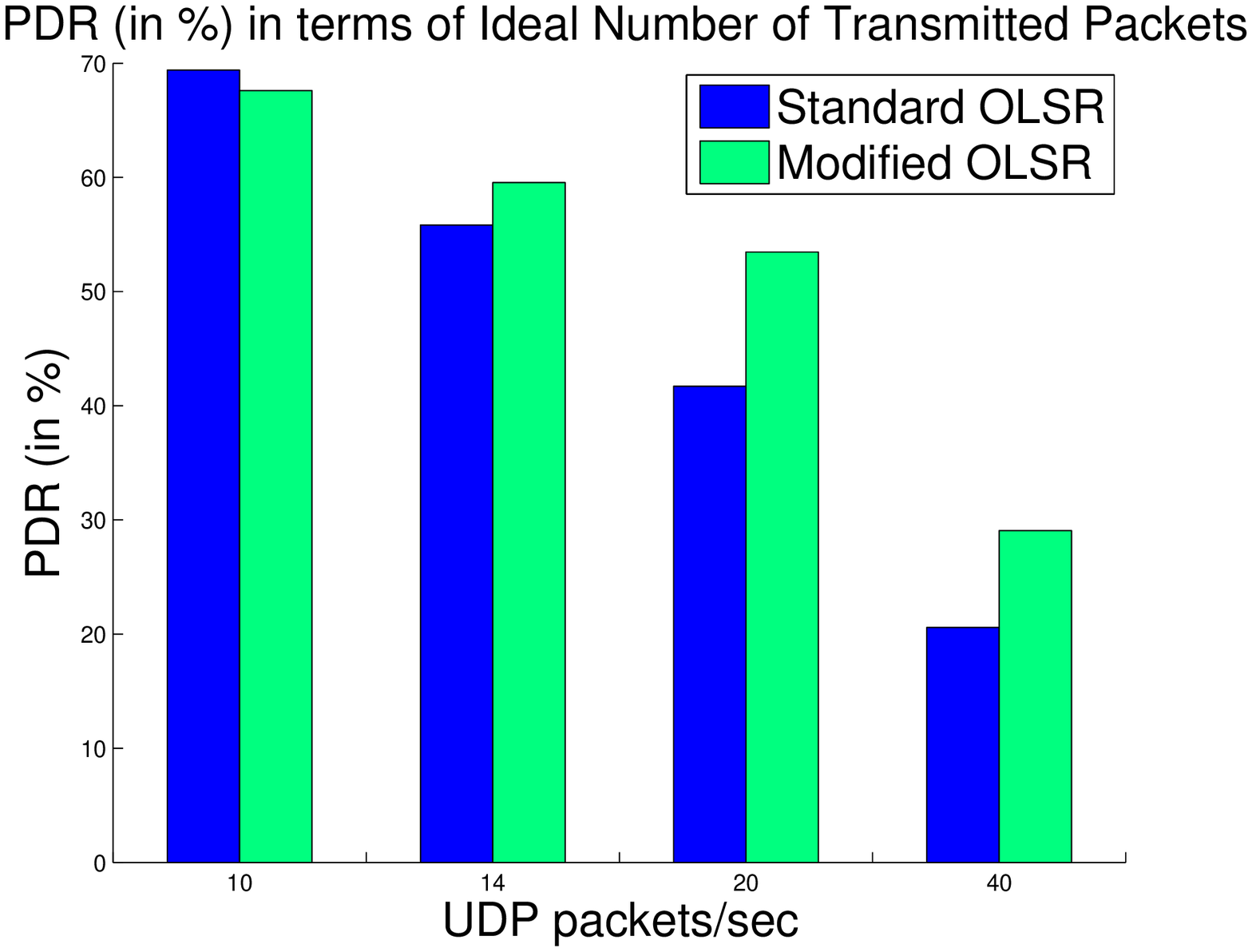}
                \caption{PDR for Various Packet Interarrival times}
                \label{fig:PDR Low Mobility}
        \end{subfigure}%
        ~ 
        \begin{subfigure}[b]{0.25\textwidth}
                \includegraphics[width=\textwidth]{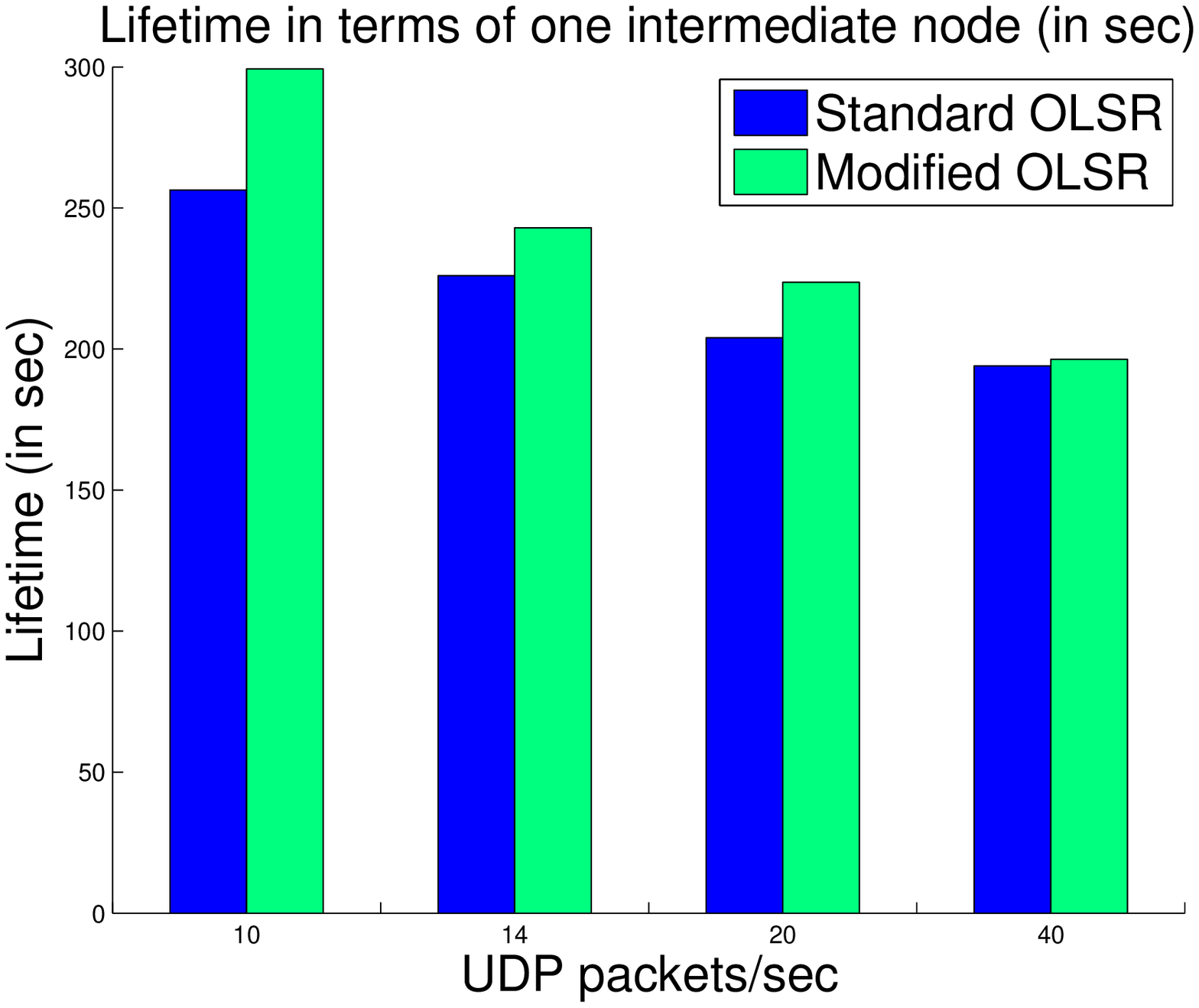}
                \caption{Network Lifetime for Vari-\\ous Packet Interarrival times}
                \label{fig:Network Lifetime Low Mobility}
        \end{subfigure}
  \caption{Low Mobility Scenario}
  \label{fig:Low Mobility Scenario}
\end{figure}

\begin{figure}
        \centering
        \begin{subfigure}[b]{0.25\textwidth}
                \includegraphics[width=\textwidth]{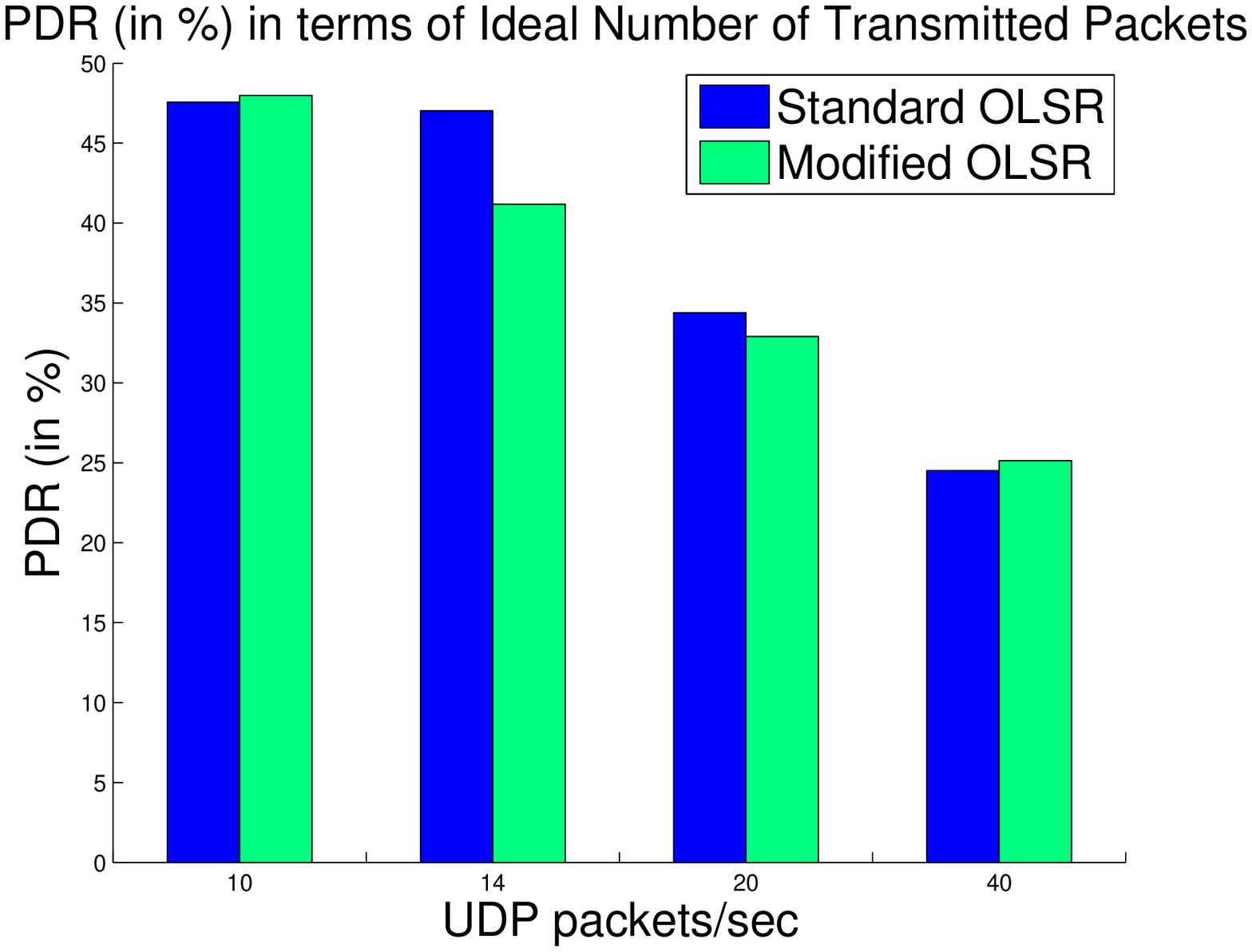}
                \caption{PDR for Various Packet Interarrival times}
                \label{fig:PDR High Mobility}
        \end{subfigure}%
        ~ 
        \begin{subfigure}[b]{0.25\textwidth}
                \includegraphics[width=\textwidth]{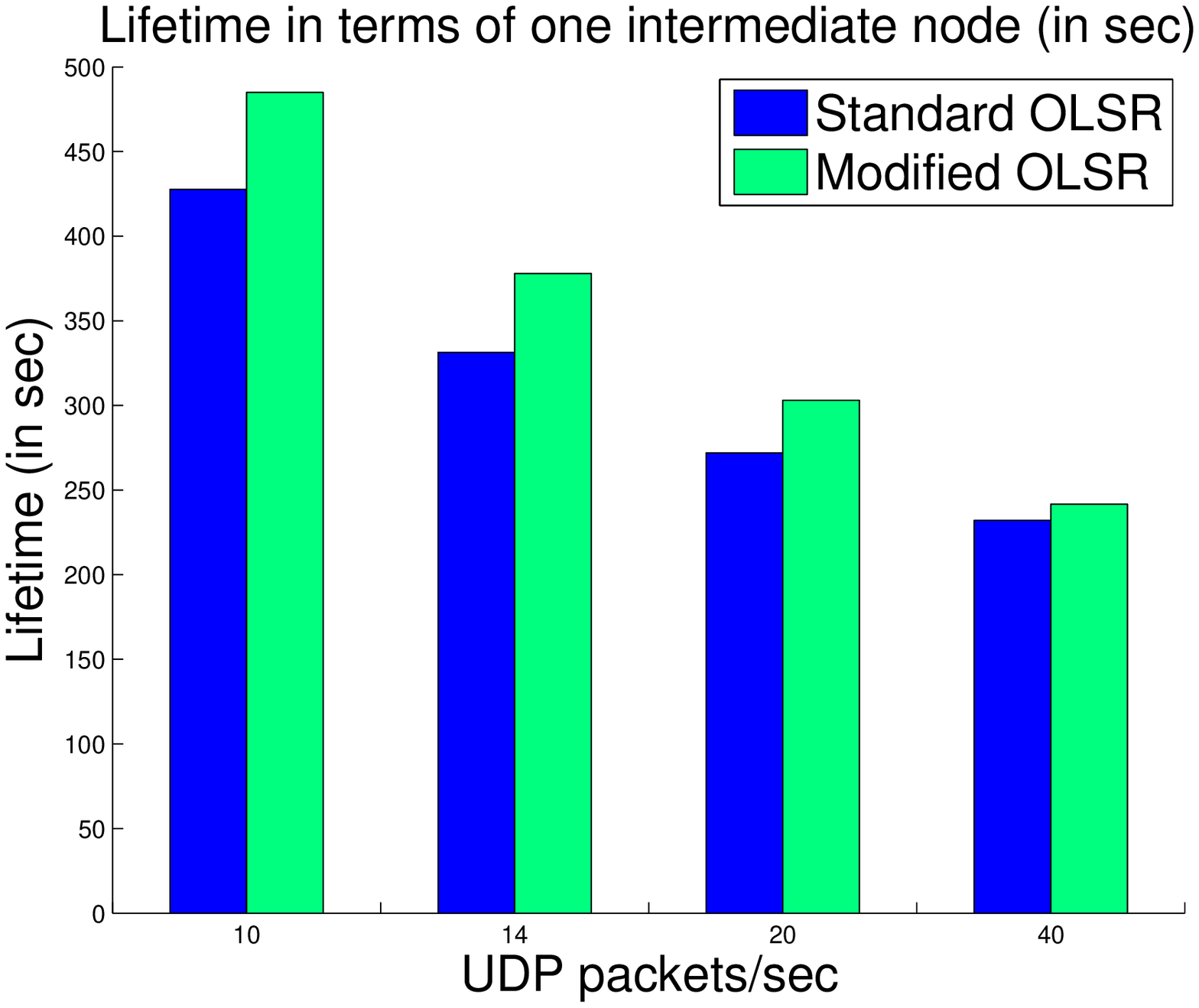}
                \caption{Network Lifetime for Vari-\\ous Packet Interarrival times}
                \label{fig:Network Lifetime High Mobility}
        \end{subfigure}
  \caption{High Mobility Scenario}
  \label{fig:High Mobility Scenario}
\end{figure}

In the low mobility scenario, we observe (Figure~\ref{fig:Low Mobility Scenario}) similar results with the static scenario. There is 5-15\% improvement in network lifetime and PDR for the reasons explained above. In the high mobility scenario, the improvement in the network lifetime is lower than in the other cases (Figure~\ref{fig:High Mobility Scenario}), due to the fact that the mobility imposes dynamic change of intermediate nodes selected for routing. In this scenario, both the standard and the modified OLSR select several different routes during simulation. In addition, PDR remains the same as in the standard OLSR and at some rates is being slightly decreased. This effect comes from the lack of robustness of our scheme to adapt to highly changing dynamic environments. In high mobility conditions, the modified OLSR cannot learn the weights fast enough to take the routing decisions and this leads to more than expected packet losses.

Overall, in static and low mobility scenarios, we notice better PDR than in high mobility scenarios, due to the fact that our protocols have slow convergence in rapidly changing dynamics. In addition, in high mobility scenarios, we achieve greater network lifetime than in the other two scenarios, because the protocols choose a lot of different alternative paths due to the dynamic changes in the network topology. Finally, in all three scenarios, as we increase the data rate we observe decrease in PDR and in network lifetime. In the case of higher rates, a lot of packets are lost due to network congestion and bandwidth limitations (link bandwidth 1 Mbps), which results to the decrease of the PDR. In addition, network lifetime is decreased due to the transmission (and retransmission from the MAC layer) of large number of packets in the case of high data rates.

\section{Conclusion and Future Work}
In this paper, we proposed a novel multi-metric energy efficient routing scheme, integrated in the OLSR routing protocol. Three cross-layer parameters, which indicate energy depletion, are considered to form a weight, representing the cost of routing through this node. Node weights are updated and sent to the network through TC packets periodically to be taken into consideration for routing decisions. We evaluated the Modified OLSR under a 
range of different scenarios, varying traffic load and 
mobility pattern. The Modified version of the OLSR achieves significant increase in network lifetime (5-20\%), without loss of performance (in terms of PDR). 

Future work includes experimenting on larger networks. In large networks we may have slow routing protocol convergence by relying to TC messages, hence the robustness of our scheme needs to be investigated. In addition, we are currently working on the formulation of an optimization problem, in order to assign appropriate weights according to the current state of the network.


\section*{Acknowledgment}

The research of E. Paraskevas and J.S. Baras was partially supported by grants NSF CNS-1018346,  
NSF CNS-1035655, and NIST 70NANB11H148.



\bibliographystyle{./IEEEtran}
\bibliography{./IEEEabrv,./references}
\nocite{*}
%



\end{document}